# INCREASING INFORMATIVITY OF THE THERMALLY STIMULATED DEPOLARIZATION METHOD

S. N. Fedosov, A. E. Sergeeva and T. A. Revenyuk

*Department of Physics, Odessa National Academy of Food Technologies, Odessa, Ukraine*

Two modifications of the Thermally Stimulated Depolarization Current method are proposed to improve resolution and sensitivity of the method by connecting either a real capacitor, or an additional resistor in series with the sample. It is shown experimentally that high sensitivity of the TSDC method with an air gap can be obtained, if the gap is substituted by the capacitor, while all advantages of the method remain in force. It has been found that in one experiment it is possible not only to measure the TSD current, but also to obtain data on the Thermally Stimulated Conductivity, if the properly selected additional resistor is periodically switched on and off.

## INTRODUCTION

Thermally stimulated depolarization current measurement is one of the important methods for identifying and characterizing relaxation processes in electrified dielectrics and electrets [1,2]. At the same time, the most commonly used modification of the method with two short-circuited electrodes attached to the sample has some disadvantages and drawbacks. In particular, depolarization currents caused by relaxation of homo- and heterocharge flow in one direction making difficult separation and characterization of the components. Moreover, it is very probable that the intrinsic resistance of the sample in the high temperature part of the experiment becomes comparable, or even smaller than that of the measuring device. As the result, the real TSD current is distorted and even stray currents are often observed. In this paper we suggest two modification of the TSDC method free of the above-mentioned drawbacks.

In order to make homo- and heterocharge relaxation currents flow in opposite directions, a modification of the TSDC method has been proposeed with an air gap introduced between the sample surface and one of the electrodes [3]. It has been proved that the air gap must be as narrow as possible to provide for the reasonable sensitivity and in any case to be much smaller than the thickness of the sample. However, in the case of thin polymer electrets having thickness from about 10 to 30 μm, it is difficult to realize and maintain such an air gap. That is why, the air gap is sometimes replaced by a thin non-polar dielectric spacer, for example a Teflon film inserted between the sample and the electrode [2,3]. In this work we suggest a much better solution for improving resolution and sensitivity of the TSDC method.

## REAL CAPACITOR REPLACES THE DIELECTRIC GAP

From electrical point of view, the gap between the electrode and the sample or the spacer works as a capacitor $C$ connected in series with the sample (Fig. 1). Hence, the expression for the depolarization current through the gap can be written as

$$I = C \frac{dU}{dt} \qquad (1)$$

with

$$C = \varepsilon_o \varepsilon \frac{A}{d_1}, \qquad (2)$$

where $U$ is the gap voltage that changes as the result of the charge accumulation at the gap boundaries, $\varepsilon_o$ the permittivity of a vacuum, $\varepsilon$ the dielectric constant of the spacer ($\varepsilon=1$ in the case of the air gap), $d_1$ the thickness of the dielectric gap, $A$ the surface area of the sample and the spacer. It is clear from Eqs. (1) and (2) that the gap must be as narrow as possible to obtain reasonable sensitivity of the method.

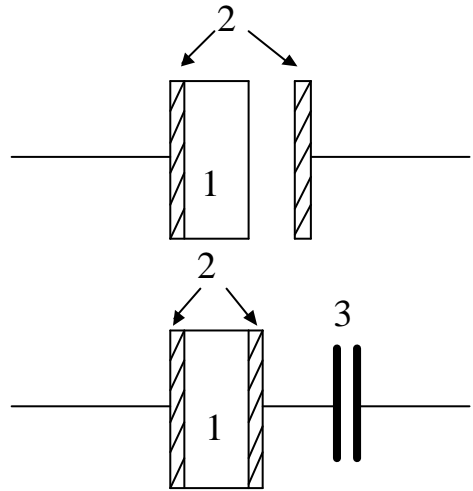

Fig. 1. Schematic diagram showing equivalence of the air gap between the sample and the electrode with the real capacitor connected in series with the sample. 1 – sample, 2 – electrodes, 3 – capacitor.

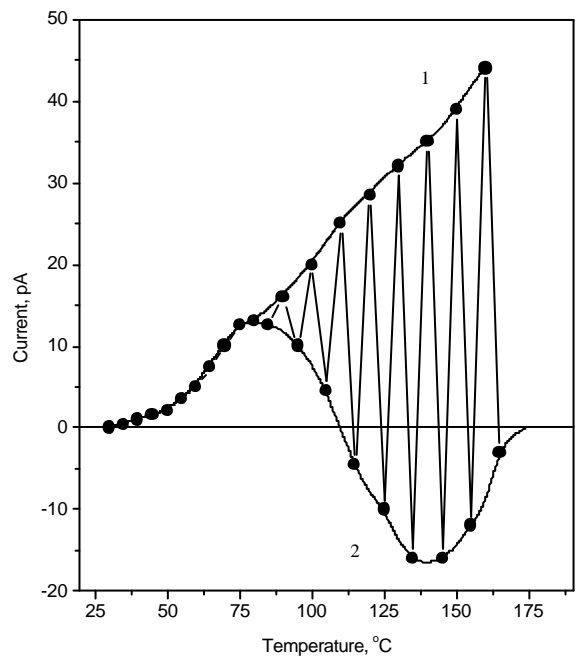

Fig. 2. Temperature dependence of the current during the TSD experiment with the two-side electroded PVDF-PZT composite sample. The series capacitor of 0.1 µF is periodically switched on (1) and off (2). The heating rate 3 K/min, the sample thickness 560 µm.

Assuming that thickness of the Teflon film spacer is 10 µm, the dielectric constant of the spacer $e=2,3$ and the surface area $A=1$ cm², one obtains $?=200$ pF, which is not enough to guarantee high sensitivity for the sample thickness in the range from 10 to 30 µm. From the other side, it is impractical to use a thinner spacer because of the breakdown problems.

We have found experimentally that good results can be obtained, if the air gap or the dielectric spacer is replaced by a real capacitor, provided that its capacitance is much higher than that of the sample. In this case, the considerable increase in sensitivity is observed comparable with the sensitivity in TSDC experiments with non-blocking short-circuited electrodes.

This is confirmed by the data in Fig. 2 showing the TSD current while the capacitor connected in series with the sample is periodically switched on and off. It is seen from Fig. 2 that at the initial stage the same TSD current is observed with and without the capacitor. Calculations show that in order to obtain the same sensitivity with a dielectric gap, its thickness must be of the order of 0.02 µm (assuming $e$=2.3) that cannot be practically realized.

With increase of temperature, as follows from Fig. 2, the difference between the TSD current with and without the capacitor increases, while, according to the theoretical model, the current at the end of the depolarization process should decrease to zero in both cases. Abrupt current increase in short-circuited samples is probably caused not by the relaxation process, but is rather a result of stray EMFs that under increased conductivity induce large parasitic currents. The suppression of the stray currents is one of the basic advantages of the proposed method.

Thus, by periodically switching the series capacitor on and off, one can obtain in one experiment the two TSDC curves corresponded to the short-circuited mode and the mode with the dielectric gap. By comparing the two curves, one can make conclusions on the nature of relaxation processes and calculate their parameters. For example, from the data presented in Fig. 2 it is clear that there are two relaxation processes in PVDF-PZT composites related to homo- and heterocharge. It is seen that the homocharge is more stable than the heterocharge. Besides, there are non-compensated stray EMFs in the experimental setup causes most probably by potential differences between contacting metals.

We took polymer-ceramics composites here only as an example for showing applicability of the proposed modification of the TSD current method. We also examined the method on pure polymer electrets, such as uniaxially stretched PVDF films of 25 µm thickness, and non-linear optical polymer films of polystyrene doped with DR1 chromophore and poled in a corona triode.

## ADDITIONAL RESISTOR IN SERIES WITH THE SAMPLE

It is known that during the measurement of a current, the intrinsic resistance of the current source $r$ must be much larger that the input resistance $R$ of the ammeter ($r>>R$), otherwise, due to redistribution of voltage between the sample and the ammeter, reading of the meter would not correspond to the real value of the current.

In the case of the TSD current measurements, the resistance of the meter remains constant, while that of the sample decreases with time and temperature. Thus even if the condition $r(T)>>R$ initially is met, it might be destroyed in the high temperature region and the current shown by the meter would become larger than the real depolarization current. Moreover, stray currents will not be limited any more by the resistance of the sample, so the meter will show continuously increasing current, while the real depolarization current should go to zero at the end of the depolarization experiment.

To avoid the complications, we propose to connect an additional resistor $R'$ in series with the sample (Fig. 3). The value of the resistor $R'$ is selected in such a way that it should be much smaller than the resistance of the sample $r$ in the low temperature range of the TSDC experiment, while to be much larger than the latter in the high temperature range, i.e. it should satisfy the following conditions

$$R'<<r(T) \text{ at low temperatures,}$$
$$R'>>r(T) \text{ at high temperatures.} \qquad (3)$$

If the above conditions are met, the TSD current will not be distorted in either temperature range.

As one can see from Fig. 4, the TSD current in the low temperature part with the additional resistor is the same as the current without the resistor, indicating that $R'$ can be neglected at low temperatures comparing to $r(T)$.

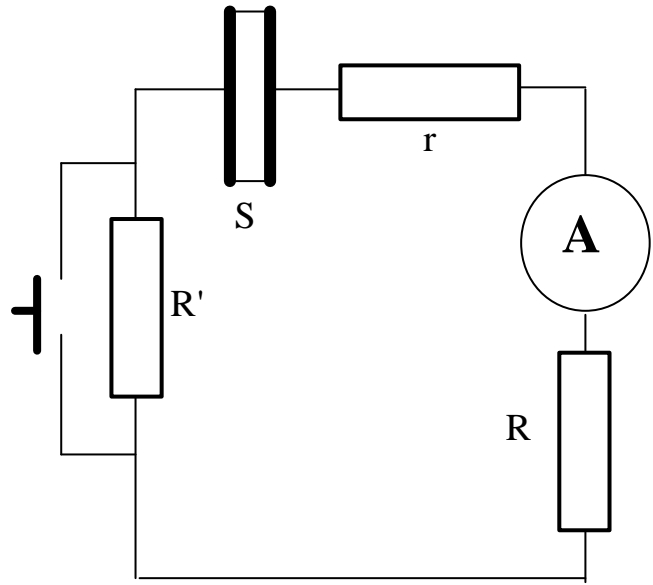

Fig. 3. Schematic diagram showing how the series resistor $R'$ is connected in the measuring circuit consisting of the sample $S$, its intrinsic resistance $r$, and the meter A with its input resistance $R$. The additional resistor $R'$ is periodically switched on and off.

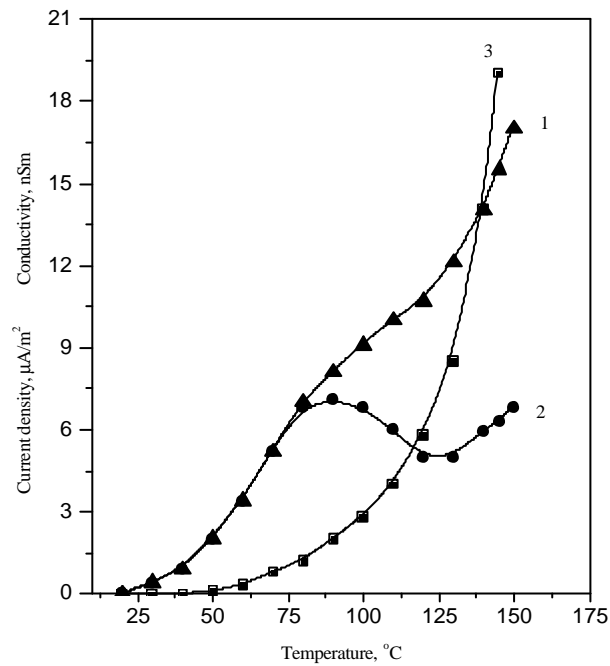

Fig. 4. TSD current curves for a charged composite PVDF-PZT obtained with the additional resistor of 220 MΩ that was periodically switched on (1) and off (2). Also shown is the temperature dependence of conductivity (3) derived from the curves (1) and (2). The electret was formed at $T=100°C$, $E=12$ MV/m during $t=0.5$ h.

In the high temperature part of the experiment the current with the additional resistor becomes much more relief than without the resistor, with relaxation peaks clearly distinguishable, because now $R'$ "substitutes" $r(T)$ that drastically decreases with temperature, while the stray currents are also suppressed.

We have found one important feature of the suggested mode of the TSD current measurements. By periodic switching on and shortening the additional resistor $R'$ one can obtain two TSD curves, from which the temperature dependence of the intrinsic specific conductivity $g(T)$ of the sample can be easily obtained. Simple calculations based on the schematic diagram shown in Fig. 3 gives the following expression

$$g(T) = \frac{d}{A} \frac{I(T) - I'(T)}{[I'(T) \cdot (R + R') - I(T) \cdot R]} \qquad (4)$$

where $I(T)$ is the TSD current without the additional resistor $R'$, $I'(T)$ the current with the resistor $R'$, $d$ the sample thickness, $A$ the surface area of the sample.

All three curves – two experimental ones and one calculated – are shown in Fig. 4. As one can see, the temperature dependence of conductivity shows a typical exponential behavior.

## CONCLUSION

Two methods for improving informativity of the TSD current measurements are proposed and its application is illustrated. In the first one, an additional capacitor is connected in series with the sample resembling the TSD mode with the air gad or the dielectric spacer, but with much higher sensitivity and with suppression of the stray currents. In the second method, a properly selected resistor is connected in series with the sample and the ammeter. In this case, if the additional resistor is switched on and short circuited during the TSD experiment, one obtains information as if the two thermally stimulated methods are combined in one experiment, namely the TSD current measurement and the TS conductivity measurement, since one gets simultaneously data on both the TSD current and the thermally stimulated conductivity (TSC). It is important that the measurement of the conductivity is performed without any external power supply, but rather under internal self-balanced electric field. In this case, the higher resolution power is expected in comparison with the traditional method.

## REFERENCES

[1] Yu. A. Gorokhovatsky, *Fundamentals of the Thermal Depolarization Analysis*, Moscow: Nauka, 1981. (Russ.)
[2] J. van Turnhout, *Thermally Stimulated Discharge of Polymer Electrets*, Amsterdam–Oxford–New York: Elsevier, I975.
[3] G. M. Sessler (ed.) *Electrets*, Vol. 1, Third Edition, Laplacian Press, Morgan Hill, 1999.